\documentclass[10pt,prl,twocolumn,superscriptaddress]{revtex4}

\usepackage{graphicx}
\usepackage{amssymb}
\usepackage{times}
\usepackage{amsmath}
\usepackage{amsfonts}
\usepackage{txfonts}
\usepackage{color}
\usepackage{natbib}
\usepackage{units}

\begin{document}

\title{Feedback Enhanced Sensitivity in Optomechanics: Surpassing the Parametric Instability Barrier}

\author{Glen I. Harris} \affiliation{Centre for Engineered Quantum Systems, University of Queensland, St Lucia, Queensland 4072, Australia}
\author{Ulrik L. Andersen} \affiliation{Department of Physics, Technical University of Denmark, Building 309, 2800 Lyngby, Denmark }
\author{Joachim Knittel} \affiliation{Centre for Engineered Quantum Systems, University of Queensland, St Lucia, Queensland 4072, Australia}
\author{Warwick P. Bowen} \affiliation{Centre for Engineered Quantum Systems, University of Queensland, St Lucia, Queensland 4072, Australia}

\begin{abstract}
The intracavity power, and hence sensitivity, of optomechanical sensors is commonly limited by parametric instability. Here we characterize the parametric instability induced sensitivity degradation in a micron scale cavity optomechanical system. Feedback via optomechanical transduction and electrical gradient force actuation is applied to suppress the parametric instability. As a result a $5.4$ fold increase in mechanical motion transduction sensitivity is achieved to a final value of $1.9\times 10^{-18}\rm m Hz^{-1/2}$.

\end{abstract}
\maketitle

\begin{figure}[!ht]
\begin{center}
\includegraphics[width=7cm]{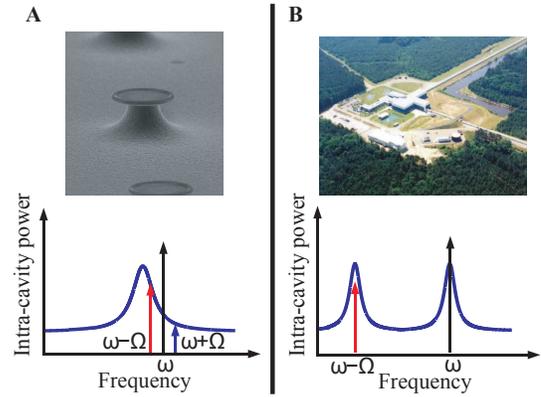}
\caption{Ultra sensitive optomechanical systems exhibit parametric instability through cavity enhancement of the Stokes sideband produced by their mechanical motion. {\bf A} In detuned micron scale optomechanical systems Stokes and anti-Stokes sidebands spectrally overlap with the same optical mode. {\bf B} In large-scale interferometers Stokes sidebands must overlap with an adjacent optical mode. \emph{Image of LIGO Livingston Laboratory courtesy of Skyview Technologies}}
\label{sidebands}
\end{center}
\end{figure}

 Optical techniques are capable of ultra-precise measurements of phase, position and refractive index, with the measurement sensitivity typically limited by optical shot noise which can be reduced by maximising the optical power. Using coherent states of light the ultimate sensitivity is fundamentally set by the Standard Quantum Limit (SQL) \cite{Caves80,Teufel09}. However, well before the SQL is reached the radiation pressure may become sufficiently strong to severely alter the dynamics of the intrinsic mechanical motion of the sensor. This regime, called parametric instability, is characterized by violent mechanical oscillations and was first theoretically investigated by Braginsky \cite{Braginsky01} in the context of large scale interferometers for gravitational wave detection followed by experimental observation in electrical readout of resonant bar systems \cite{Cuthbertson96} and later in optical micro-cavities \cite{Rokhsari06}. The physical process, described graphically in Fig.\ref{sidebands}, is a result of radiation pressure from asymmetric Stokes and anti-Stokes sidebands generated from the mechanical motion of the cavity. If this process, known as dynamical backaction heating \cite{Kippenberg07}, amplifies the motion at a rate faster than the mechanical decay rate then parametric instability occurs. Due to a combination of large mechanical oscillations and necessary saturation of amplification, the noise floor of the optomechanical sensor increases, rendering it ineffective at transducing small signals. Parametric instability is predicted to be a potential problem in the context of the advanced Laser Interferometer Gravitational Observatory (LIGO) \cite{Braginsky01, Ju09} and, more generally, in many cavity optomechanical systems designed for ultra-precise sensing.

Parallel to the development of ultra-precise large scale optical sensors for gravitational wave detection, there has been a recent push towards real time read out and control of mesoscopic mechanical oscillators in the quantum regime \cite{Schwab05,Oconnell10,Teufel11_Cooling}. Their classical counterparts are extensively used in applications ranging from chronometry to ultra fast sensors and actuators ~\cite{Ekinci05,Oconnell10}. While quantum mechanical oscillators promise to enhance applications in sensing and metrology, perhaps the most exciting prospects lie in fundamental research where such systems could enable new quantum information technologies \cite{Mancini03}, experimental tests of quantum nonlinear mechanics \cite{Woolley08, Lifshitz08, Rugar91} and even quantum gravity \cite{Marshall03}. A stringent requirement is that transduction capabilities be limited only by unavoidable quantum noise sources, enabling measurements at the standard quantum limit (SQL) of the oscillator. However, in many situations this is precluded by parametric instability occurring not only when detuned to the blue (heating) side of the optical resonance but also with zero detuning \cite{Ivanov95}. Furthermore both the generation of non-classical mechanical states and pushing below the SQL require techniques such as back-action evasion (BAE) \cite{Schwab10} which are typically limited by instabilities.

In this paper we model a feedback scheme designed to eliminate parametric instability, revealing a simple transfer function that is insensitive to fluctuations of many experimental parameters. We experimentally implement the feedback scheme and characterize the degradation in, and feedback induced revival of, sensitivity due to parametric instability. Our cavity opto-electromechanical system (COEMS), seen in Fig. \ref{sidebands}, consists of a silica microtoroid integrating high $Q$ mechanical and optical modes with strong electrical actuation~\cite{Lee10}. Parametric instability is found to occur at optical powers of $60\mu \rm W$ resulting in a drastic loss in sensitivity for higher power levels, and a maximum optomechanical sensitivity still a factor of twenty higher than the SQL. Upon application of a viscous dampening force via electric feedback, stabilization of the parametric instability was achieved, allowing sensitivities at the level of $1.9\times 10^{-18}\rm m Hz^{-1/2}$, limited by available optical power.

The dynamical interaction between light and mechanical motion including radiation pressure $F_{\rm rad}$, feedback $F_{\rm fb}$ and thermal forces $F_{\rm T}$ can be described through the equations of motion \cite{Kippenberg07}
\begin{eqnarray}
&m\left[\ddot{x}+ \Gamma_0 \dot{x} +  \omega^2_m x\right] = F_{\rm rad} + F_{\rm T} + F_{\rm fb}  \\
	&\dot{a} = -\left[\gamma -i(\Delta_{0}+g x)\right] a + \sqrt{2\gamma_{in}}a_{in}
\end{eqnarray}
The first equation describes the motion of the mechanical oscillator where $m$, $\Gamma_0$ and $\omega_m$ are its effective mass, damping rate and resonance frequency, respectively; $F_{\rm rad} = \hbar g |a(t)|^{2}$, and $F_{\rm T}=\sqrt{ \Gamma_0 k_{B} T m} \xi(t)$ where $\xi(t)$ is a unit white noise Wiener process. The second equation describes the intra-cavity optical field where $\Delta_{0}$ is the optical detuning, $|a|^2$ is the intra cavity photon number and $|a_{in}|^2$ is the input photon flux, coupled into the cavity at rate $\gamma_{in}$. The total optical decay rate is $\gamma = \gamma_{in}+\gamma_{0}$ where $\gamma_{0}$ is the intrinsic decay rate. The equations are coupled via the optomechanical coupling parameter, $g$, which gives rise to both static and dynamic effects such as radiation pressure bistability \cite{Dorsel83}, the optical spring effect \cite{Sheard04}, and dynamical backaction cooling and amplification \cite{Arcizet06,Gigan06,Kippenberg05}. Due to the nonlinear nature of the equations of motion linearization is required to reach an analytic solution where a separation of each variable into its mean value and flucuations is performed; $a=\bar{a}+\delta a$ and $x = \bar{x}+\delta x$. Taking the linearized equations into the frequency domain yields
\begin{eqnarray}
\delta a(\omega) \!&=&\! \frac{\sqrt{2\gamma_{in}}\delta \tilde{a}_{in}+ i g \bar{a}\delta x(\omega)}{\gamma -i\left(\Delta-\omega\right)}\label{eq:daf} \\ 
 \chi_{0}^{-1}\delta x(\omega) \!&=&\! \hbar g \left[\bar{a}\delta a^{\dagger}(-\omega)+ \bar{a}^{\ast}\delta a(\omega)\right]+\! F_{\rm T}(\omega)\! +\! F_{\rm fb}(\omega) \label{eq:dxf}
\end{eqnarray}
where $\chi_0 =m^{-1}\left[ \omega^2_m -\omega^2+i\Gamma_0\omega \right]^{-1}$ is the mechanical susceptibility and $\Delta = \Delta_0 +g \bar{x}$ is the static detuning of the cavity in the presence of radiation pressure. As seen in Eq.~(\ref{eq:daf}) the mechanical fluctuations are imprinted onto the field $\delta a$ which, in turn, is out-coupled and detected on a photodiode giving a photocurrent $i = a^{\dagger}_{\rm out} a_{\rm out}$. After some work the resulting photocurrent fluctuation, $\delta i(\omega) = \bar{a}^{\ast}_{\rm out} \delta a_{\rm out}+ \bar{a}_{\rm out} \delta a^{\dagger}_{\rm out}$, is found to be
\begin{eqnarray}
\delta i = \delta x \left(\frac{2ig|\bar{a}|^2\Delta \left[\omega - i2\gamma_{0} \right] }{ \gamma + \Delta-\omega^{2} +i2\gamma \omega} \right)+ \delta i_{a}
\label{eq:photocurrent}
\end{eqnarray}
where $\delta i_{a}$ contains all noise terms associated with the input field. This signal is then applied back onto the oscillator via the feedback force $F_{\rm fb}(\omega) = G\delta i$ where $G$ is a complex feedback gain. Substituting this feedback force and the optical fluctuations, given by Eq.~(\ref{eq:daf}), into Eq.~(\ref{eq:dxf}), an analytic form can be obtained for the modification of mechanical motion due to radiation pressure and feedback forces combined. The terms contributing to $\delta x$ modify the mechanical susceptibility, $\chi$, such that 
\begin{eqnarray}
\chi^{-1} = \chi_{0}^{-1} + \frac{2 g|\bar{a}|^2 \Delta\left[\hbar g + G\left(i\omega -2\gamma_{0}\right)\right]}{\gamma^2+\Delta^2-\omega^2+ i2\gamma\omega},
\label{eq:chi'}
\end{eqnarray}
If no feedback is applied, corresponding to $G = 0$, the mechanical susceptibility is modified purely by radiation pressure \cite{Kippenberg05}. As is well known the phase of the modulating radiation pressure depends on the sign of the detuning $\Delta$ resulting in either mechanical linewidth narrowing or broadening. Parametric instability occurs when the modified mechanical linewidth is negative, with correspondingly exponential amplification of the mechanical oscillations. This amplification process eventually saturates to give a steady-state linewidth $\Gamma_{\rm ss}$ close to zero but positive. Similar to mode competition in a laser this limits the parametric instability to one mechanical mode. Due to the large shifts of the optical resonance from amplified mechanical motion the average intra-cavity power, and hence radiation pressure, decreases resulting in saturation of the mechanical amplification. Since the mechanism for transduction is equivalent to actuation, the nonlinear saturation comes together with nonlinear transduction. This nonlinearity acts to mix different frequency components in the spectrum resulting in broadband noise and hence severely degrading the transduction sensitivity. From Eq.~(\ref{eq:chi'}) it can be seen that this degradation can be canceled and the original mechanical susceptibility recovered if the gain is chosen to be
\begin{eqnarray}
G_{\rm crit} = \frac{\hbar g} {2\gamma_0-i\omega},
\label{eq:Gain}
\end{eqnarray}
such that all modifications to the mechanical susceptibility from radiation pressure are canceled by the electrical feedback. This simple expression is completely insensitive to flucuations in the detuning $\Delta$, the coupling rate $\gamma_{\rm in}$ and the input power $|a_{\rm in}|^2$, making the feedback system very robust against external noise sources. This robustness necessarily translates to simplicity of implementation, which is all important for optomechanical systems pushing towards the SQL. Many techniques have been proposed for the stabilization of parametric instabilities such as the addition of acoustic dampers, thermal control and active feedback from a transduction signal using optical, electrical or mechanical actuation~\cite{Ju09}; but feedback stabilization has to date been demonstrated only in large scale low frequency systems~\cite{Corbitt06,Blair95} where the parametric instability occurs due to the presence of many optical modes (Fig.\ref{sidebands}B) rather than in one optical mode as is the case here (Fig.\ref{sidebands}A), and with no characterisation of the degradation in sensitivity due to parametric instability, or the enhancement achieved via active feedback.
 
\begin{figure}[ht]
\begin{center}
\includegraphics[width=7cm]{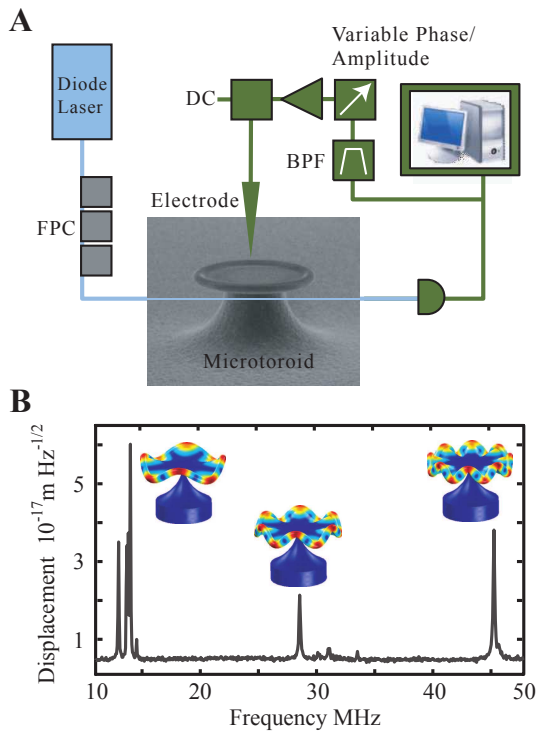}
\caption{{\bf A} Experimental schematic. Blue (light grey) indicates the optical components allowing transduction and parametric instability; Green (dark gray) indicate the electrical components used in the feedback stabilization. FPC: Fiber polarization controller. BPF: Bandpass filter. {\bf B} Observed mechanical spectra below parametric instability threshold}
\label{ExpSch}
\end{center}
\end{figure}

Our experimental setup to demonstrate feedback suppression of parametric instability in a micron-scale system is shown in Fig.\ref{ExpSch}. A tunable diode laser at $780\rm nm$ was evanescently coupled into a microtoroidal whispering gallery mode using a tapered optical fiber. The microtoroid had major and minor diameters of $60\mu \rm m$ and $6\mu \rm m$ respectively with a $25\mu \rm m $ undercut. The toroid-taper separation was controlled by a piezo stage to allow critical coupling into the optical cavity. The laser was thermo-optically locked \cite{Carmon04, McRae09} to the full width half maximum (FWHM) of the optical mode which had an intrinsic quality factor of $Q\approx 10^{7}$. This optical detuning allowed simultaneous radiation pressure induced mechanical amplification and transduction of the mechanical motion. The absolute mechanical displacement amplitude was calibrated via the optical response to a known reference phase modulation~\cite{Schliesser08}. The mechanical motion, which modulates the optical resonance frequency, was detected via fluctuations in the transmitted power incident on an InGaAs photodiode. Fourier analysis of the photocurrent reveals a mechanical power spectra with peaks corresponding to mechanical resonances. A typical spectra containing many mechanical modes, and their corresponding finite element simulation, can be seen in Fig.~\ref{ExpSch}B at optical powers below the threshold for parametric instability. At higher optical powers the $4^{\rm th}$ order crown mode at $14\rm MHz$ experiences parametric instability, degrading the sensitivity with which other mechanical modes may be transduced, which will subsequently be stabilized by the feedback loop. Here we focus on the $6^{\rm th}$ order crown mode at $28.6\rm MHz$, which was characterized experimentally to have an effective mass and linewidth of $0.3\mu \rm g$ and $90\rm kHz$ respectively, resulting in a standard quantum limit of $S_{SQL}^{1/2} = \sqrt{ \frac{\hbar}{2 m_{eff} \Omega_{0} \Gamma }} = 8\times10^{-21} \rm m Hz^{-1/2}$. 
\begin{figure}[ht!]
\begin{center}
\includegraphics[width=7cm]{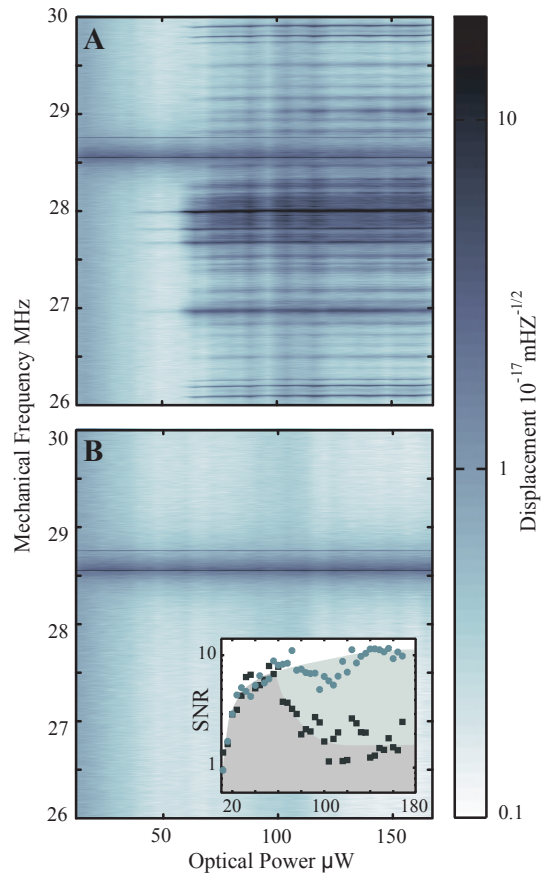}
\caption{Mechanical power spectra {\bf B} and {\bf A} show, respectively, the mechanical spectra with and without feedback stabilization of the regenerative $14\rm MHz$ mode. Inset: Signal to noise ratio of the $28.6 \rm MHz$ mechanical mode versus optical power with feedback, circles, and without, squares.
}
\label{pcolour}
\end{center}
\end{figure}

\begin{figure}[ht]
\begin{center}
\includegraphics[width=7cm]{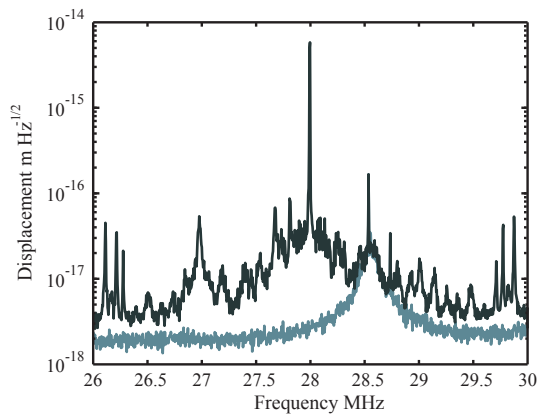}
\caption{ Mechanical spectra observed with $160\mu \rm W$ of input power, well above the parametric instability threshold, with and without feedback stabilization of the unstable $14 \rm MHz$ mode.}
\label{MechSpectra}
\end{center}
\end{figure}

We see from Fig.~\ref{pcolour}A that at low power the $6^{\rm th}$ order crown mode is easily resolved with the sensitivity improving with increasing optical power as expected from photon shot noise statistics. However, as the input power is increased above $60\mu W$ the $4^{\rm th}$ order crown mode experiences parametric instability, generating harmonics on the transduction signal at $28, 42$ and $56 \rm MHz$ due to the nonlinear process involved in saturation. This is evident by the emergence of a dark narrow band, amongst broadband noise, at $28\rm MHz$ in Fig.~\ref{pcolour}A. Added noise at a fixed power of $160\mu\rm W$ is shown in Fig.~\ref{MechSpectra} (dark line) and reveals the narrowed harmonic at $28\rm MHz$, and a group of three peaks equally spaced on either side which are due to mixing with other mechanical modes near the unstable mode. In addition to these harmonics and beats, $\frac{1}{f}$ noise, thermo-refractive noise and signals from other mechanical modes are also mixed with the $14\rm MHz$ signal causing broadband noise. This extra noise is clearly illustrated in Fig.~\ref{MechSpectra} (dark trace) where it completely obscures the motion of the $6^{\rm th}$ order crown mode (grey trace). The SNR of the $6^{\rm th}$ order crown mode is shown as a function of power in the inset to Fig.~\ref{pcolour}B (black squares), with severe degradation apparent once threshold is reached. 

Feedback to suppress the parametric instability was implemented by electrically filtering and amplifying the photocurrent and applying it directly to a sharp electrode placed close to the microtoroid. This facilitated strong electrical actuation of the mechanical motion through electrical gradient forces \cite{McRae10}. Consecutive bandpass filters were used to isolate the unstable mechanical mode at $14 \rm MHz$ allowing maximum amplification of the feedback signal while minimizing the effect of feedback on nearby mechanical modes. The feedback phase and gain were controlled inside the feedback loop by an electronically variable phase shifter and attenuator respectively. With correct gain and phase as given in Eq.~\ref{eq:Gain}, the viscous damping force applied by feedback fully suppressed the parametric instability and eliminated the harmonics and associated noise from the unstable $14\rm MHz$ mode, as shown in Fig.~\ref{pcolour}B and Fig.~\ref{MechSpectra}A (light line). Consequently, the transduction sensitivity of the $6^{\rm th}$ order crown mode was found to improve with optical power, even above threshold, as shown in the inset to Fig.~\ref{pcolour}B (cirles).

This work is an important step to reaching the standard quantum limit in micron-sized cavity optomechanical systems designed for ultra-precise sensing and, more generally, for suppressing parametric instabilities in systems involving BAE.

This research was funded by the Australian Research Council Centre of Excellence CE110001013 and Discovery Project DP0987146. Device fabrication was undertaken within the Queensland Node of the Australian Nanofabrication Facility. We gratefully acknowledge The Danish Council for Independent Research (Sapere Aude program).

\end{document}